\documentclass[preprint,10pt,a4paper,twocolumn,final,tightenlines,showpacs,bibnotes,nofootinbib,notitlepage,pra]{revtex4}
\usepackage{amssymb}

\usepackage{amsmath}

\newtheorem{theorem}{Theorem}
\newtheorem{acknowledgement}[theorem]{Acknowledgement}

\input{tcilatex}

\begin{document}

\title{Decoherence of Josephson charge qubit}
\author{Xian-Ting Liang\thanks{%
Electronic address: xtliang@ustc.edu}}
\affiliation{Department of Physics and Institute of Modern Physics, Ningbo University,
Ningbo, 315211, China}

\begin{abstract}
In this paper we investigate decoherence time of superconducting Josephson
charge qubit (JCQ). Two kinds of methods, iterative tensor multiplication
(ITM) method derived from the qusiadiabatic propagator path integral (QUAPI)
and Bloch equations method are used. Using the non-Markovian ITM method we
correct the decoherence time predicted by Bloch equations method. By
comparing the exact theoretical result with the experimental data we suggest
that the Ohmic noise plays the central role to the decoherence of the JCQ.

\end{abstract}

\maketitle

\section{Introduction}

Solid state qubits are considered promising candidates for making processors
of quantum computers because they can be scaled up to a large numbers. The
qubits based on Josephson junction are these kinds of qubits. But how about
their other qualities, in particular, how about their coherence? Many
efforts not only theoretical \cite{Decoherence-Therotical,Martinisetal} but
also experimental \cite{nature1999,charge-qubit2,flux-qubit2,phase-qubit1}
have been contributed to search decoherence time as well as decoherence
mechanism of the qubit systems. The theoretical researches are in general
based on the spin-boson model \cite{Weiss, Leggett}. By now, it is suggested
that there are $1/f$ noise and Ohmic noise in the environment of the
Josephson qubits. It is considered that the former is derived from the
background charge fluctuations and the latter from the electromagnetic noise
due to voltage fluctuations. But what is the most primary mechanism of the
decoherence, or which is the mainly noise source in the environment of the
qubits? By using the Bloch equations \cite{Bloch-Equation, Makhlinetal} one
can derive the relaxation time and dephasing time of the qubits. However, in
the derivation of Bloch equations an approximation scheme in general the
Markov approximation should be used. It has been pointed out recently, the
Markov approximation is not a suitable approximation scheme in the
investigations of the quantum system for quantum computing purposes because
it is not usually valid at low temperatures and for short cycle time of
quantum computation \cite{Privman}. A similar viewpoint on a qubit of double
quantum dots is also pointed out, see as \cite{cond-mat/0505621}. Privman
and his co-worker \cite{Privman-coworkers} investigated the decoherence of
qubits with short-time approximation rather than the Markov one, and many
interesting and novel results are obtained. However, it is not enough to
only investigate the decoherence in a short time because the coherence in a
longer time for some qubits (for example qubits for quantum registers) is
also interesting. Fortunately, it is found out that by using short-time
propagators one can construct a path integral \cite{Feynman} in a longer
time. The well established iterative tensor multiplication (ITM) algorithm
derived from the qusiadiabatic propagator path integral (QUAPI) \cite{Makri}
can be used to solve the evolutions of low-dimension open quantum systems in
a moderate long time \cite{Thorwartetal}. In this method the temporal
non-local interactions is involved and it is non-Markovian. Thus, we expect
that the non-Markovian method can be used to investigate the decoherence of
Josephson qubits and then to detect the mainly mechanism of the decoherence
in the qubits. There are many kinds of Josephson qubit models \cite%
{charge-qubit1,charge-qubit2,flux-qubit1,flux-qubit2,phase-qubit1,phase-qubit2,science2001,Youetal}%
, but in this paper we only investigate the Josephson charge qubit (JCQ). We
shall obtain an accurate decoherence time of the JCQ by using the ITM
algorithm without the Markovian approximation. Base on the accurate
decoherence time we shall suggest a mainly mechanism of the decoherence in
the JCQ.

\section{Models}

The elementary unit of the quantum computer is a qubit which is in fact a
two-level quantum system \cite{Nielsen-Chuang}. There are many physical
realizations for the system. However, any single qubit can be represented as
a spin-1/2 particle, and its Hamiltonian can be written as $H\left( t\right)
=-\frac{1}{2}\vec{B}\left( t\right) \cdot \hat{\sigma}.$ Here $\sigma
_{x,y,z}$ are Pauli matrixes in a space of states $\left| \uparrow
\right\rangle =\left( 
\begin{tabular}{ll}
0 & 1%
\end{tabular}%
\right) ^{T}$ and $\left| \downarrow \right\rangle =\left( 
\begin{tabular}{ll}
1 & 0%
\end{tabular}%
\right) ^{T}$ which form basis states of the Hamiltonian. The quantity $\vec{%
B}\left( t\right) $ has different physical meaning according to the
difference of the physical realizations of the qubits. For example, if the
qubit is realized by a spin of some particle the $\vec{B}\left( t\right) $
will be an effective magnetic field. But to the JCQ, the components of the
``magnetic field'' are set as $B_{x}=E_{J},$ where $E_{J}$ is the Josephson
energy of the Josephson junction, $B_{y}=0,$ and $B_{z}=4E_{C}\left(
1-2n_{g}\right) $ \cite{Makhlinetal}. The Hamiltonian of a general qubit can
be represented as 
\begin{equation}
H_{s}=-\frac{1}{2}B_{z}\sigma _{z}-\frac{1}{2}B_{x}\sigma _{x}.  \label{e1}
\end{equation}%
If one modulates the gate voltage and makes $n_{g}=1/2,$ the JCQ system may
be reduced to $H_{s}=-\frac{1}{2}B_{x}\sigma _{x}$. In this paper, we take $%
E_{J}=51.8$ $\mu eV,$ and $E_{C}=122$ $\mu eV$ according to Ref. \cite%
{nature1999}$.$ If one considers the interaction of the qubit and its
environment, and takes the environment as a bath, the whole Hamiltonian of
the system-bath will be \cite{Makhlinetal} 
\begin{equation}
H=H_{0}+H_{env},  \label{e2}
\end{equation}%
where%
\begin{eqnarray}
H_{0} &=&H_{s},  \notag \\
H_{env} &=&\tsum\nolimits_{k}\left[ \frac{1}{2m_{k}}p_{k}^{2}+\frac{1}{2}%
m_{k}\omega _{k}^{2}\left( x_{k}-\frac{\lambda _{k}}{m_{k}\omega _{k}^{2}}%
\sigma _{z}\right) ^{2}\right] .  \notag \\
&&  \label{e3}
\end{eqnarray}%
In general, $H_{0}$ should plus counterterms $-\tsum\nolimits_{k}\lambda
_{k}^{2}/2m_{k}^{2}\omega _{k}^{4}$ which can ensure that some important
features of the qubit do not depend on the coupling strength. In our
problem, the counterterms only contribute a global phase so we can ignore
it. This is the well-known spin-boson model, a appropriately reduced open
qubit model. In the following, we firstly analyze this model. On the one
hand, for the bath, only the linearly coupling term is chosen in the
coordinates $x_{k}$, representing the lowest nontrivial term in the Taylor
series expansion of the potential. It is accurate enough in the weak
coupling case. On the other hand, for the qubit, only $\sigma _{z}$ coupling
term is included. The terms of $\sigma _{x,y}$ coupling with the bath have
not been included in the Hamiltonian. The reason is that $\sigma _{x,y}$
have only off-diagonal matrix elements in the $\sigma _{z}$ representation,
i.e., they only change $\left| \uparrow \right\rangle $ to $\left|
\downarrow \right\rangle $ and vice versa.

In order to obtain the reduced density matrix of the qubit in qubit-bath
system, one should know the coupling coefficients $\lambda _{k}$ in Eq.(\ref%
{e3}). However, we do not need to know their details because all
characteristics of the bath pertaining to the dynamics of the observable
system are captured in the spectral density function of noise%
\begin{equation}
J_{X}\left( \omega \right) =\frac{\pi }{2}\sum_{k}\frac{\lambda _{k}^{2}}{%
m_{k}\omega _{k}^{2}}\delta \left( \omega -\omega _{k}\right) .  \label{e4}
\end{equation}%
In the case of truly macroscopic environment the spectral density is for all
practical purposes a continuous function of frequency. In the following we
shall see that the spectral density function $J\left( \omega \right) $
instead of $\lambda _{k}$ is directly used in obtaining the elements of the
reduced density matrix. It is related to the power spectrum of the noise as%
\begin{equation}
S_{X}\left( \omega \right) =J_{X}\left( \omega \right) \hbar \coth \left(
\omega \beta \hbar /2\right) .  \label{e5}
\end{equation}%
Here, $\beta =1/k_{B}T,$ where $k_{B}$ is the Boltzmann constant and $T$ is
the temperature. Throughout the paper we take $T=30$ mK according to Ref. %
\cite{nature1999}. Due to very wide of the real and imaginary parts of the
response function (see Eq.(\ref{e14}) in the following) in the time range
for the $1/f$ bath, we in fact cannot investigate the evolutions of the
reduced density matrix of qubit in the $1/f$ bath with the ITM. Thus, in
this paper, we focus on the case that the environment is the Ohmic bath. The
spectral density of the Ohmic bath can be expressed as \cite{Makhlinetal}%
\begin{equation}
J_{X}\left( \omega \right) =2\pi \hbar \alpha \omega \exp \left( -\frac{%
\omega }{\omega _{C}}\right) ,  \label{e6}
\end{equation}%
where$\ \omega _{C}$ is the cut-off frequency of the bath modes. The
parameter $\alpha $ is dimensionless strength of the dissipation which is
determined by concrete qubit-bath system. For the JCQ model, Makhlin \emph{%
et al.} proposed a numerical simulation value $\alpha \approx 10^{-6}$. In
this paper we suppose $\alpha =5\times 10^{-6}.$ If $\alpha >5\times 10^{-6}$
the decoherence time will be shorter than the results obtained in this
paper. On the other hand, if $\alpha <5\times 10^{-6}$ the decoherence time
will be longer than the results in this paper.

\section{QUAPI and ITM}

In the following, we firstly review the QUAPI method and then the ITM \cite%
{Makri} scheme. Suppose the initial state of the qubit-bath system has the
form%
\begin{equation}
R\left( 0\right) =\rho \left( 0\right) \otimes \rho _{bath}\left( 0\right) ,
\label{e7}
\end{equation}%
where $\rho \left( 0\right) $ and $\rho _{bath}\left( 0\right) $ are initial
states of the qubit and bath. The evolution of its reduced density operator
of the open qubit%
\begin{equation}
\tilde{\rho}\left( s^{\prime \prime },s^{\prime };t\right) =\text{Tr}%
_{bath}\left\langle s^{\prime \prime }\right| e^{-iHt/\hbar }\rho \left(
0\right) \otimes \rho _{bath}\left( 0\right) e^{iHt/\hbar }\left| s^{\prime
}\right\rangle  \label{e8}
\end{equation}%
is given by%
\begin{eqnarray}
&&\tilde{\rho}\left( s^{\prime \prime },s^{\prime };t\right)  \notag \\
&=&\sum_{s_{0}^{+}=\pm 1}\sum_{s_{1}^{+}=\pm 1}\cdot \cdot \cdot
\sum_{s_{N-1}^{+}=\pm 1}\sum_{s_{0}^{-}=\pm 1}\sum_{s_{1}^{-}=\pm 1}\cdot
\cdot \cdot \sum_{s_{N-1}^{-}=\pm 1}  \notag \\
&&\times \left\langle s^{\prime \prime }\right| e^{-iH_{0}\Delta t/\hbar
}\left| s_{N-1}^{+}\right\rangle \cdot \cdot \cdot \left\langle
s_{1}^{+}\right| e^{-iH_{0}\Delta t/\hbar }\left| s_{0}^{+}\right\rangle 
\notag \\
&&\times \left\langle s_{0}^{+}\right| \rho \left( 0\right) \left|
s_{0}^{-}\right\rangle  \notag \\
&&\times \left\langle s_{0}^{-}\right| e^{iH_{0}\Delta t/\hbar }\left|
s_{1}^{-}\right\rangle \cdot \cdot \cdot \left\langle s_{N-1}^{-}\right|
e^{iH_{0}\Delta t/\hbar }\left| s^{\prime }\right\rangle  \notag \\
&&\times I\left( s_{0}^{+},s_{1}^{+},\cdot \cdot \cdot
,s_{N-1}^{+},s^{\prime \prime },s_{0}^{-},s_{1}^{-},\cdot \cdot \cdot
,s_{N-1}^{-},s^{\prime };\Delta t\right) ,  \notag \\
&&  \label{e9}
\end{eqnarray}%
where the influence functional is%
\begin{eqnarray}
&&I\left( s_{0}^{+},s_{1}^{+},\cdot \cdot \cdot ,s_{N-1}^{+},s^{\prime
\prime },s_{0}^{-},s_{1}^{-},\cdot \cdot \cdot ,s_{N-1}^{-},s^{\prime
};\Delta t\right)  \notag \\
&=&\text{Tr}_{bath}\left[ e^{-iH_{env}\left( s^{\prime \prime }\right)
\Delta t/2\hbar }e^{-iH_{env}\left( s_{N-1}^{+}\right) \Delta t/2\hbar
}\right.  \notag \\
&&\times \cdot \cdot \cdot e^{-iH_{env}\left( s_{0}^{+}\right) \Delta
t/2\hbar }\rho _{bath}\left( 0\right) e^{iH_{env}\left( s_{0}^{-}\right)
\Delta t/2\hbar }  \notag \\
&&\left. \times \cdot \cdot \cdot e^{iH_{env}\left( s_{N-1}^{-}\right)
\Delta t/2\hbar }e^{iH_{env}\left( s\prime \right) \Delta t/2\hbar }\right] .
\label{e10}
\end{eqnarray}%
The discrete path integral representation of the qubit density matrix
contains temporal nonlocal terms $I\left( s_{0}^{+},s_{1}^{+},\cdot \cdot
\cdot ,s_{N-1}^{+},s^{\prime \prime },s_{0}^{-},s_{1}^{-},\cdot \cdot \cdot
,s_{N-1}^{-},s^{\prime };\Delta t\right) $ which denotes the process being
non-Markovian. With the quasiadiabatic discretization of the path integral,
the influence functional, Eq.(\ref{e10}) takes the form%
\begin{equation}
I=\exp \left\{ -\frac{i}{\hbar }\tsum_{k=0}^{N}\tsum_{k^{\prime
}=0}^{k}\left( s_{k}^{+}-s_{k}^{-}\right) \left( \eta _{kk^{\prime
}}s_{k^{\prime }}^{+}-\eta _{kk^{\prime }}^{\ast }s_{k^{\prime }}^{-}\right)
\right\} ,  \label{e11}
\end{equation}%
where $s_{N}^{+}=s^{\prime \prime }$ and $s_{N}^{-}=s^{\prime }.$ The
coefficients $\eta _{kk^{\prime }}$ can be obtained by substituting the
discrete path into the Feynman-Vernon expression. Their expressions have
been shown in \cite{Makri}. Thus, the influence functional can be expressed
with a product of terms corresponding to different $\Delta k$ as 
\begin{eqnarray}
I &=&\tprod_{k=0}^{N}I_{0}\left( s_{k}^{\pm }\right)
\tprod_{k=0}^{N-1}I_{1}\left( s_{k}^{\pm },s_{k+1}^{\pm }\right)
\tprod_{k=0}^{N-\Delta k}I_{\Delta k}\left( s_{k}^{\pm },s_{k+\Delta k}^{\pm
}\right)  \notag \\
&&...\tprod_{k=0}^{N-\Delta k_{\max }}I_{\Delta k_{\max }}\left( s_{k}^{\pm
},s_{k+\Delta k_{\max }}^{\pm }\right) .  \label{e12}
\end{eqnarray}%
Here, $\Delta k=k-k^{\prime },$ where $k^{\prime }$ and $k$ are points of
discrete path integral expressions, see Ref.\cite{Makri}, and%
\begin{eqnarray}
I_{0}\left( s_{i}^{\pm }\right) &=&\exp \left\{ -\frac{1}{\hbar }\left(
s_{i}^{+}-s_{i}^{-}\right) \left( \eta _{ii}s_{i}^{+}-\eta _{ii}^{\ast
}s_{i}^{-}\right) \right\} ,  \notag \\
I_{\Delta k}\left( s_{i}^{\pm },s_{i+\Delta k}^{\pm }\right) &=&\exp \left\{
-\frac{1}{\hbar }\left( s_{i+\Delta k}^{+}-s_{i+\Delta k}^{-}\right) \right.
\notag \\
&&\times \left. \left( \eta _{i+\Delta k,i}s_{i}^{+}-\eta _{i+\Delta
k,i}^{\ast }s_{i}^{-}\right) \right\} ,\Delta k\geqslant 1.  \notag \\
&&  \label{e13}
\end{eqnarray}%
The length of the memory of the time can be estimated by the following bath
response function%
\begin{equation}
\gamma \left( t\right) =\frac{1}{\pi }\int_{0}^{\infty }d\omega J\left(
\omega \right) \left[ \coth \left( \frac{\beta \hbar \omega }{2}\right) \cos
\omega t-i\sin \omega t\right] .  \label{e14}
\end{equation}%
It is shown that when the real and imaginary parts behave as the delta
function $\delta \left( t\right) $ and its derivative $\delta ^{\prime
}\left( t\right) ,$ the dynamics of the reduced density matrix is Markovian.
However, if the real and imaginary parts are broader than the delta function
the dynamics is non-Markovian. The broader of the Re$[\gamma \left( t\right)
]$ and Im$[\gamma \left( t\right) ]$ are, the longer of the memory time will
be. The broader of the Re$[\gamma \left( t\right) ]$ and Im$[\gamma \left(
t\right) ]$ are, the more serious the Markov approximation will distort the
practical dynamics. In Fig.1 we plot Re$[\gamma \left( t\right) ]$ and Im$%
[\gamma \left( t\right) ]$ of the Ohmic bath.%
\begin{eqnarray*}
&& \\
&& \\
&&Fig.1 \\
&& \\
&&
\end{eqnarray*}%
From Fig. 1 we see that the memory time is about $\tau _{mem}=1\times
10^{-11}$ $s$ for the Ohmic bath. Due to nonlocality of the influence
functional, it is impossible to calculate the reduced density matrix by Eq.(%
\ref{e9}) in matrix multiplication scheme. However, the short range
nonlocality of the influence functional Eq.(\ref{e10}) implies that the
affects of the nonlocality should drop off rapidly as the ``interaction
distance''\ increases. In the Makri's ITM scheme the interaction can be
taken into account at each iteration step. The reduced density matrix at
time $t=N\Delta t$ ($N$ even) is given as%
\begin{equation}
\tilde{\rho}\left( s_{N}^{\pm },N\Delta t\right) =A^{\left( 1\right) }\left(
s_{N}^{\pm };N\Delta t\right) I_{0}\left( s_{N}^{\pm }\right) ,  \label{e15}
\end{equation}%
where%
\begin{eqnarray}
A^{\left( 1\right) }\left( s_{k+1}^{\pm };(k+1)\Delta t\right) &=&\int
ds_{k}^{\pm }T^{\left( 2\right) }\left( s_{k}^{\pm },s_{k+1}^{\pm }\right)
\label{e16} \\
&&\times A^{\left( 1\right) }\left( s_{k}^{\pm };k\Delta t\right) .  \notag
\end{eqnarray}%
Here,%
\begin{eqnarray}
&&T^{\left( 2\Delta k_{\max }\right) }\left( s_{k}^{\pm },s_{k+1}^{\pm
}...s_{k+2\Delta k_{\max }-1}^{\pm }\right)  \notag \\
&=&\tprod_{n=k}^{k+\Delta k_{\max }-1}K\left( s_{k}^{\pm },s_{k+1}^{\pm
}\right) I_{0}\left( s_{n}^{\pm }\right) I_{1}\left( s_{n}^{\pm
},s_{n+1}^{\pm }\right)  \notag \\
&&\times I_{2}\left( s_{n}^{\pm },s_{n+2}^{\pm }\right) ...I_{\Delta k_{\max
}}\left( s_{n}^{\pm },s_{n+\Delta k_{\max }}^{\pm }\right) ,  \label{e17}
\end{eqnarray}%
and%
\begin{equation}
A^{\left( \Delta k_{\max }\right) }\left( s_{0}^{\pm },s_{1}^{\pm
},...,s_{\Delta k_{\max }-1}^{\pm };0\right) =\left\langle s_{0}^{+}\right|
\rho _{s}\left( 0\right) \left| s_{0}^{-}\right\rangle ,  \label{e18}
\end{equation}%
where%
\begin{eqnarray}
K\left( s_{k}^{\pm },s_{k+1}^{\pm }\right) &=&\left\langle
s_{k+1}^{+}\right| \exp (-iH_{0}\Delta t/\hbar )\left| s_{k}^{+}\right\rangle
\notag \\
&&\times \left\langle s_{k}^{-}\right| \exp (iH_{0}\Delta t/\hbar )\left|
s_{k+1}^{-}\right\rangle .  \label{e19}
\end{eqnarray}%
In the ITM scheme a short-time approximation instead of the Markov
approximation is used. The approximation makes a error of short-time
propagator in order $\left( \Delta t\right) ^{3}$ which is small enough as
we set the time step $\Delta t$ very small. It is shown that when the time
step $\Delta t$ is not larger than the characteristic time of the qubit
system which can be calculated with $\hbar /E_{J}$ the calculation is
accurate enough \cite{Privman}. In particular, the scheme do not discard the
memory of the temporal evolution, which may appropriate to solve the
decoherence of qubits.

\section{Decoherence of Josephson qubits}

To measure effects of decoherence one can use the entropy, the first
entropy, and many other measures, such as maximal deviation norm etc., see %
\cite{Privman}. However, essentially, the decoherence of a open quantum
system is reflected through the decays of the off-diagonal coherent terms of
its reduced density matrix. The decoherence is in general produced due to
the interaction of the quantum system with other systems with a large number
of degrees of freedom, for example the devices of measurement or
environment. In this paper, we investigate the decoherence time of the JCQ
via directly describing the evolution of the off-diagonal coherent terms,
instead of using any measure of decoherence. In our following
investigations, we suppose the cut-off frequency of the bath modes is $%
\omega _{C}=5$ $\left( ps\right) ^{-1}$. We set the initial state of the
qubit $\rho \left( 0\right) =\frac{1}{2}\left( \left| 0\right\rangle +\left|
1\right\rangle \right) \left( \left\langle 0\right| +\left\langle 1\right|
\right) $ which is a pure state and it has the maximum coherent terms, and
the initial state of the environment $\rho _{bath}\left( 0\right)
=\prod\nolimits_{k}e^{-\beta M_{k}}/$Tr$_{k}\left( e^{-\beta M_{k}}\right) .$

\emph{Decoherence time obtained from ITM scheme:} At first, we use the ITM
scheme investigating the decoherence time of the Josephson qubits. The
evolutions of the coherent elements of the reduced density matrix of the JCQ
in the Ohmic bath is plotted in Fig.2. Here, we simply choose $\Delta
k_{\max }=1$ and $\Delta t=1.27\times 10^{-11}$ s in the ITM scheme. The
choice of the time step is feasible as we consider that it should be not
shorter than the memory time of the bath, because the latter is about $\tau
_{mem}=1\times 10^{-11}$ s for the Ohmic bath, see Fig. 1. It is also
appropriate as we consider that the time step should be not longer than the
characteristic time of the qubit, where the latter is about $\tau =1.3\times
10^{-11}$ s.%
\begin{eqnarray*}
&& \\
&& \\
&&Fig.2 \\
&& \\
&&
\end{eqnarray*}%
It is shown that when we choose the parameter of the dimensionless strength
of the dissipation $\alpha =5\times 10^{-6},$ the decoherence time of the
JCQ is about $\tau _{2}=1.05299$ $\mu $s.

\emph{Decoherence time calculated on Bloch equations: }It is well known that
the decoherence time can be derived on Bloch equations. In this method, the
relaxation and dephasing times $\tau _{1},$ $\tau _{2}$ can be calculated as %
\cite{BBS}%
\begin{equation}
\tau _{1}^{-1}=2\tau _{2}^{-1}=\frac{1}{2\hbar }J\left( \omega _{0}\right)
\coth \left( \beta \hbar \omega _{0}/2\right) ,  \label{e20}
\end{equation}%
where $\omega _{0}=B_{x}/\hbar $ is the natural frequency of the Josephson
qubit. From Eq.(\ref{e20}) and using the same parameters of the qubit and
the bath as above we can obtain that the decoherence time is $\tau
_{2}=1.61966$ $\mu $s$.$ It is shown that the time obtained from Eq.(\ref%
{e20}) is longer than that from the ITM scheme. We suggest that the
difference is derived from the following two reasons. The first is that the
Bloch equations are in general derived from the Markov approximation which
discards the memory of the bath in the derivation of the dynamical
evolution. The second is that the Eq.(\ref{e20}) is obtained from the second
order approximation of perturbation series. The decoherence of the qubit
described with this method is only the ``resonant decoherence'' \cite{Openov}%
. It is not equals to the actual decoherence accurately except for the
``nonresonant decoherence'' very small.

\emph{Compared with the experimental results: }In our calculations, we use
the parameters similar to Ref.\cite{nature1999}, so we can compare our
results to the experimental decoherence time. In \cite{nature1999} the
decoherence time of a single-Cooper pair box, namely, the JCQ is estimated.
The main decoherence source is thought to be spontaneous photon emission to
the electromagnetic environment (which can just be described by the Ohmic
bath). In \cite{nature1999} the authors pointed out that the experimental
decoherence time of the JCQ could exceed $1$ $\mu $s$.$ It is shown that by
use of the Ohmic decoherence mechanism we can obtain a theoretical
decoherence time of the JCQ not only by the ITM scheme but also through the
Bloch equations method. Both of the theoretical results are agreement with
the experimental one very well!

\section{Conclusions}

In this paper we investigated the decoherence time of the JCQ in the Ohmic
bath with the ITM scheme based on the QUAPI and based on the Bloch
equations. The results derived from the two kinds of methods are compared
with each other. It is shown that the decoherence time obtained from the
Bloch equations method is longer than that from the ITM scheme. We suggest
that the difference is resulting from the different choices of the
approximation scheme because the Markov approximation used in the Bloch
equations method discards the memory of the bath. It is also because the
Bloch equations method discards the higher order decoherence, namely, only
the ``resonant decoherence'' \cite{Openov} is left over. So the decoherence
time obtained from this method is not equals to the actual decoherence time
accurately. What is more important to us is that the experimental
decoherence time of the JCQ due to spontaneous photon emission is well
agreement with the ITM decoherence time of the JCQ because of the
electromagnetic fluctuations. Both of the spontaneous photon emission and
the electromagnetic fluctuations have the same decoherence mechanism and can
be modeled by the Ohmic bath. These can lead to a conclusion that the Ohmic
bath decoherence is a central mechanism in the JCQ and the decoherence time
of the JCQ is about $1$ $\mu $s when the temperature is about $30$ mK and
the Josephson energy is about $51.8$ $\mu $ev. The decoherence time is
decided by the decoherence mechanism and affected by the experimental
temperature. If the experimental temperature increase the dimensionless
strength of the dissipation $\alpha $ will also increase and the decoherence
time will be shorter, and vice versa.

\begin{acknowledgement}
The project was supported by National Natural Science Foundation of China
(Grant No. 10347133) and Ningbo Youth Foundation (Grant No. 2004A620003).
\end{acknowledgement}

\section{Captions of the figures}

Fig.1: Real (line) and imaginary (short line) part of the response function
of the Ohmic bath. Here, we set the temperature $T=30$ mK, $\alpha =5\times
10^{-6},$ and the unit of time is second (s).

Fig.2: The evolution of the off-diagonal elements of the reduced density
matrix for the JCQ in the Ohmic bath. Here, we set $B_{x}=51.8$ $\mu $eV$,$ $%
B_{z}=0,$ $T=30$ mK, $\omega _{C}=5$ $\left( ps\right) ^{-1}$ Hz, $\alpha
=5\times 10^{-6}$, and the unit of time is picosecond (ps). The initial
state of the qubit and environment see the text.


\begin{thebibliography}{99}
\bibitem{Decoherence-Therotical} Y. Makhlin, G. Sch\"{o}n, and A. Shnirman,
Rev. Mod. Phys. 73, 357 (2001).

\bibitem{Martinisetal} J. M. Martinis, S. Nam, J. Aumentado, and K. M. Lang,
Phy. Rev. B 67, 094510 (2003).

\bibitem{Weiss} U. Weiss, \emph{Quantum Dissipative Systems}, 2nd ed.,
(World Scientific Publishing, Singapore, 1999).

\bibitem{Leggett} A. J. Leggett, S. Chakravarty, A. T. Dorsey, M. P. A.
Fisher, A. Garg and W. Zwerger, Rev. Mod. Phys. 59, 1 (1987).

\bibitem{Bloch-Equation} F. Bloch, Phys. Rev. 70, 460 (1964); R. K.
Wangsness, and F. Bloch, Phys. Rev. 89, 728 (1953); F. Bloch, Phys. Rev.
105, 1206 (1957).

\bibitem{Makhlinetal} Y. Makhlin, G. Sch\"{o}n, and A. Shnirman, Chem. Phys.
296\textbf{,} 315 (2003); New Directions in Mesoscopic Physics (Towards
Nanoscience), pp. 197-224. Eds. R. Fazio, V. F. Gantmakher, and Y. Imry,
Kluwer, Dordrecht, 2003; e-print cond-mat/0309049; A. Shnirman, Y. Makhlin,
G. Sch\"{o}n, Physica Scripta T102, 147-154 (2002); e-print cond-mat/0202518.

\bibitem{Privman} A. Fedorov, L. Fedichkin and V. Privman, J. Comp. Theor.
Nanosci. 1, 132-143 (2004); e-print cond-mat/0401248.

\bibitem{cond-mat/0505621} M. Thorwart, J. Eckel, and E. R. Mucciolo, Phys.
Rev. B. 72, 235320 (2005); e-print cond-mat/0505621. X. T. Liang, Phys. Rev.
B 72, 245328 (2005); e-print quant-ph/0506044.

\bibitem{Privman-coworkers} D. Tolkunov and V. Privman, Phys. Rev. A 69,
062309 (2004); L. Fedichkin, A. Fedorov and V. Privman, Phys. Lett. A 328,
87 (2004); V. Privman, J. Stat. Phys. 110, 957 (2003); L. Fedichkin, A.
Fedorov, Phys. Rev. A 69\textbf{,} 032311 (2004).

\bibitem{Feynman} R. P. Feynman, Rev. Mod. Phys. 20, 367 (1948); R. P.
Feynman, and A. R. Hibbs, \emph{Quantum mechanics and path integrals}
(McGraw-Hill, New York, 1965).

\bibitem{Makri} D. E. Makarov and N. Makri, Chem. Phys. Lett. 221, 482
(1994); N. Makri, J. Math. Phys. 36, 2430 (1995); N. Makri, and D. E.
Makarov, J. Chem. Phys. 102, 4600 (1995); 102, 4611 (1995).

\bibitem{Thorwartetal} M. Thorwart, P. Reimann, P. Jung, and R. F. Fox,
Chem. Phys. 235, 61 (1998); M. Thorwart, P. Reimann, and P. H\"{a}nggi,
Phys. Rev. E 62, 5808 (2000).

\bibitem{charge-qubit1} A. Shnirman, G. Sch\"{o}n, and Z. Hermon, Phys. Rev.
Lett. 79, 2371 (1997).

\bibitem{charge-qubit2} Yu. A. Pashkin, T. Yamamoto, O. Astafiev, Y.
Nakamura, D. V. Averin, and J. S. Tsai, Nature (London) 421, 823 (2003).

\bibitem{flux-qubit1} I. Chiorescu, Y. Nakamura, C. J. P. M. Harmans, and J.
E. Mooij, Science 299, (2003) 1869.

\bibitem{flux-qubit2} Y. Makhlin, G. Sch\"{o}n, and A. Shnirman, Nature
(London), 386, 305 (1999).

\bibitem{phase-qubit1} Y. Yu, S. Han, X. Chu, S.-I. Chu, and Z. Wang,
Science 296, 889 (2002).

\bibitem{phase-qubit2} M. Steffen, J. M. Martinis, and I. L. Chuang, Phys.
Rev. B 68, 224518 (2003).

\bibitem{science2001} S. Han, Y. Yu, X. Chu, S.-I. Chu, and Z. Wang, Science
293, 1457 (2001).

\bibitem{Youetal} J. Q. You, J. S. Tsai, and Franco Nori, Phys. Rev. Lett.
89, 197902 (2002); J. Q. You, J. S. Tsai, and Franco Nori, Phys. Rev. B 68,
024510 (2003); J. Q. You, Y. Nakamura, and Franco Nori, Phys. Rev. B 71,
024532 (2005).

\bibitem{Nielsen-Chuang} M. A. Nielsen and I. L. Chuang, \textit{Quantum
Computation and Quantum Information}, (Cambridge University Press,
Cambridge, England, 2000).

\bibitem{nature1999} Y. Nakamura, Y. A. Pashkin, and J. S. Tsai, Nature
(London) 398, 786 (1999).

\bibitem{BBS} B. B. Laird, J. Budimir, and J. L. Skinner, J. Chem. Phys. 94,
4391 (1991).

\bibitem{Openov} L. A. Openov, Phys. Rev. Lett. 93, 158901 (2005); E. M.
Chudnovsky, Phys. Rev. Lett. 92, 120405 (2004).
\end{thebibliography}
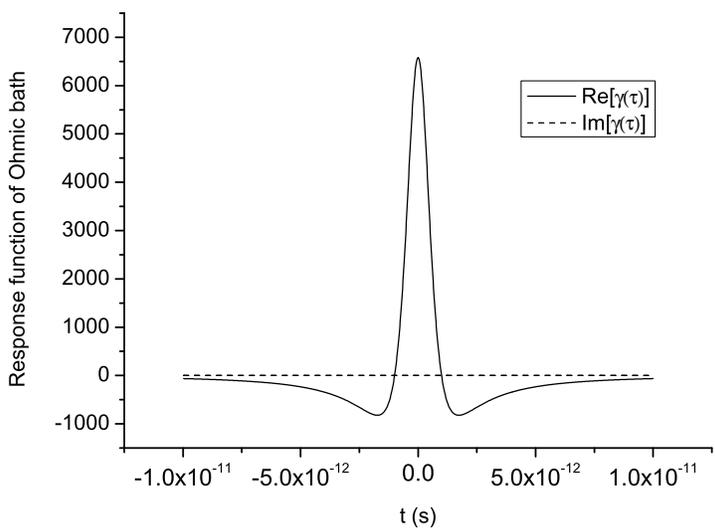
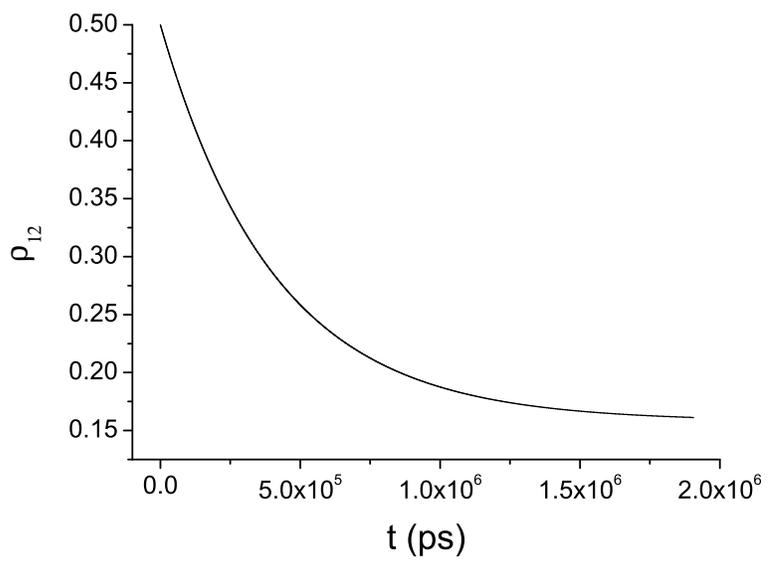
\end{document}